\documentclass[useAMS,usenatbib]{mn2e}
\usepackage{graphics,longtable}
\usepackage{graphicx}
\usepackage{amsfonts}
\usepackage{txfonts}
\usepackage{natbib}
\usepackage{amssymb}

\title[IC 2944 / 2948 complex]
      {A deep and wide-field view at the  IC 2944 / 2948 complex in Centaurus
       \thanks{Based on observations collected at ESO La Silla, under program
              076.D-0396}}
\author[Baume et al.]
  {G. Baume$^{1,2}$,
   M. J. Rodr\'{\i}guez$^{2}$,
   M. A. Corti$^{1,3}$,
   G. Carraro$^{4}$, and
   J. A. Panei$^{1,2}$
  \thanks{E-mail: gbaume@fcaglp.unlp.edu.ar (GB),
  jimenaro@fcaglp.unlp.edu.ar (MJR), mariela@fcaglp.unlp.edu.ar (MAC),
  gcarraro@eso.org (GC), panei@fcaglp.unlp.edu.ar (JAP)}\\
  $^{1}$Facultad de Ciencias Astron\'omicas y Geof\'{\i}sicas - Universidad Nacional de La Plata,
       Paseo del Bosque S/N, La Plata, (B1900FWA), Argentina\\
  $^{2}$Instituto de Astrof\'{\i}sica de La Plata (CONICET-UNLP),
       Paseo del Bosque S/N, La Plata, (B1900FWA), Argentina\\
  $^{3}$Instituto Argentino de Radioastronom\'{\i}a (CONICET),
       Cno. Gral. Belgrano Km 40 (Parque Pereyra Iraola), Berazategui, Argentina\\
  $^{4}$ESO, Alonso de C\'ordova 3107, Vitacura, Santiago de Chile, Chile}

\begin{document}

\date{Accepted 1988 December 15. Received 1988 December 14; in original form 1988 October 11}

\pagerange{\pageref{firstpage}--\pageref{lastpage}} \pubyear{2002}

\maketitle

\label{firstpage}

\begin{abstract}
We employed the ESO MPI wide-field camera and obtained deep images in the $VI_C$ pass-bands in the region of the IC~2944/2948 complex ($l \sim 294\fdg8$; $b \sim -1\fdg6$), and  complemented them with literature and archival data. We used this material to derive the photometric, spectroscopic and kinematic properties of the brightest ($V < 16$) stars in the region. The $VI$ deep photometry on the other end, helped us to unravel  the lower main sequence of a few, possibly physical,  star groups in the area.

Our analysis confirmed previous suggestions that the extinction toward this line of sight follows the normal law ($R_V = 3.1$). We could recognize B-type stars spread  in distance from  a few hundred  pc to at least 2~kpc. We found two young groups (age $\sim$ 3~Myr) located respectively at about 2.3 and 3.2~kpc from the Sun. They are characterized by a significant variable extinction ($E_{B-V}$ ranging from 0.28 to 0.45 mag), and host a significant pre-main sequence population. We  computed the initial mass functions for these groups and obtained slopes $\Gamma$  from -0.94 to -1.02 ($e_{\Gamma}$ = 0.3), in a scale where the classical Salpeter law is -1.35. We estimated the total mass of both main stellar groups in $\sim 1100$ and $\sim 500$ M$_\odot$, respectively. Our kinematic analysis indicated that both groups of stars deviate from the standard rotation curve of the Milky Way, in line with literature results for this specific Galactic direction.

Finally, along the same line of sight we identified  a third group of early-type stars located at $\sim$ 8~kpc from the Sun. This group might be located in the far side of the Sagittarius-Carina spiral arm.
\end{abstract}


\section{Introduction} \label{sec:intro}

The study of young galactic clusters and HII regions is an essential tool to improve our understanding of key astrophysical processes like star formation, its modes, spatial variation, and duration (Lada \& Lada 2003). At the same time, young star aggregates help us to probe the Galactic spiral structure, being one of the most prominent spiral arm tracers, together with gas, in the form of HI, HII, or CO (Carraro 2013).

Our group has focused its attention in recent years to several directions in the fourth quadrant of the Milky Way --see e.g. Shorlin~1 \citep{Carraro_Costa_2009}, Danks~1 and 2 \citep{Baume_et_al_2009}, NGC~6193/6167 \citep{Baume_et_al_2011} or CenOB1/NGC~4755 \citep{Corti_Orellana_2013}. In this paper we  extend our study to the remarkable IC~2944/2948 region, located at $l \sim 294\fdg8$ and $b \sim -1\fdg6$.

\begin{figure*}
\begin{tabular}{cc}
\includegraphics[width=8cm]{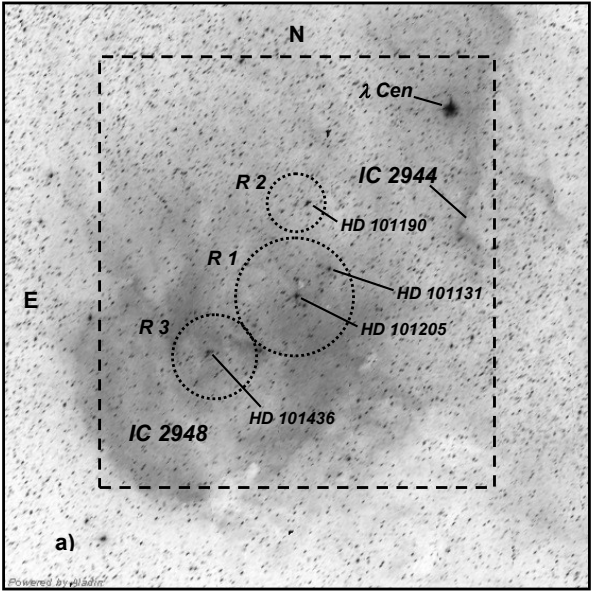}&\includegraphics[width=8cm]{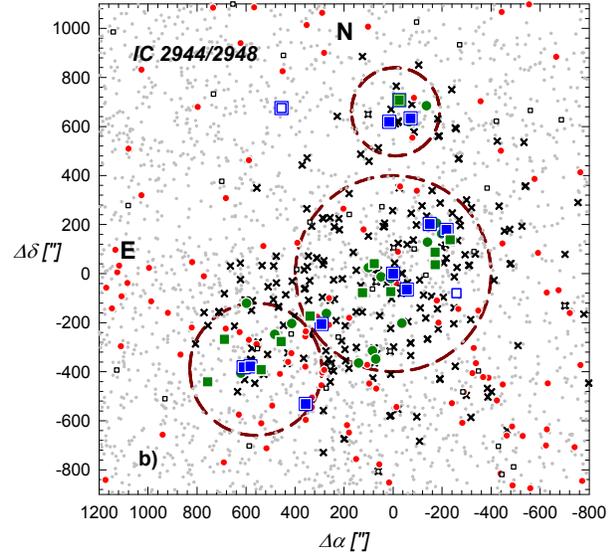}\\
\end{tabular}
\caption{a) An $ALADIN$ processed image from the Digitized Sky Survey (DSS). It is centered at HD~101205 ($\alpha_{J2000} = 11:38:20.4$; $\delta_{J2000} = -63:22:22.0$) and covering a field of $66\farcm0 \times 66\farcm0$. IC~2944/2948 HII regions and main O-type stars are identified. Big dashed square indicates the full studied area covered by our $VI_C$ data ($44\farcm6 \times 48\farcm7$) whereas the dotted circles indicate the adopted {\it selected regions} (see text). b) Schematic central part of left panel showing stars ($V<21$) as symbols. Squares and circles represent stars with and without known spectral classification respectively. Fill symbols indicate stars adopted as main two groups members (see text); blue and green symbols are O and B stars respectively. Double symbols indicate binary stars. X symbols and red dots are, respectively, x-ray sources \citep{Naze_et_al_2013} and MSX sources \citep{Lumsden_et_al_2002} correlated with our data.}
\label{fig:dss}
\end{figure*}

IC~2944 is a nebula associated to the cluster Collinder~249 (Cr~249, C1134-627,  or ``$\lambda$ Cen cluster'') and it is centered on the star HD~101205. However, the designation IC~2944 is routinely used to identify the stellar group itself (see, e.g., $SIMBAD$\footnote{http://simbad.u-strasbg.fr/simbad/},
WEBDA\footnote{http://www.univie.ac.at/webda/} ). The entire complex hosts the HII nebula IC~2948 and the stellar cluster Cr~249. The bright star $\lambda$~Cen is also located in our covered field (see Fig.~\ref{fig:dss}). According to \citet{Alter_et_al_1970}, the entire group of brightest stars in this region is also referred as Cen~OB2 association.

IC~2944/2948 appears then as a concentration of O stars (several of them identified as binary systems) and early B stars. Apparently, some of them are responsible for exciting the nebula. This area also contains Bok globules, that indicates the presence of an extremely young stellar population \citep{Reipurth_2008}.

The fundamental parameters of the various stellar groups have been matter of lively discussion in the literature, which prevented a general consensus on the star formation mode active in the region. \citet{Thackeray_Wesselink_1965} presented photoelectric photometry in the Johnson $UBV$ system and determined radial velocities for 24~stars. They found a binary fraction greater than 50\%. By using these 24 objects (19 stars located within $6'$ of the HD~101205 star and 5~O-type stars located outside this circle), they computed the following parameters: $\overline{E_{B-V}} = 0.33\pm0.06$, $V_0-M_V= 11.5 \pm 0.2$ ($\sim$ 2~kpc) and a mean visual absorption of $0\fm50$/kpc in this Galactic direction. On the other hand, \citet{Perry_Landolt_1986} studied several stars in this region using information in $V$, $uvby$ and $H_{\beta}$ bands, and claimed the stars to be a mere sample of early type field stars spread along the line of sight. \\

There is also poor agreement on the detailed structure of the complex, and the nature of the various apparent groups:

\begin{itemize}
 \item \citet{Ardeberg_Maurice_1980,Ardeberg_Maurice_1981} proposed that OB stars belong to various stellar groups  at different distances along the line of sight. Specifically, they identify three groups: a) the closest  at about 700~pc, b) seven bright stars that seem connected with HII region RWC62 at 1.7~kpc, and c) four more distant stars at 3-4~kpc.
 \item \citet{Walborn_1987} suggests that the O-type stars in the field of the HII region IC~2944 constitute a  physical cluster. This group of O stars is the cause of the  ionisation  of the surrounding region.
 \item Later, \citet{McSwain_Gies_2005} argue for the existence of a cluster, based on $b$, $y$ and $H_{\alpha}$ photometry. They estimated a distance of 1.8~kpc, an age of 6.6~Myr, and a reddening   of $E_{B-V}\sim 0.32$.
 \item \citet{Reipurth_et_al_1997} studied the globules discovered by Thackeray using CO observations  and they revealed that the two larger globules are
 kinematically separated, and have masses of 4 and 11 M$_\odot$. However, they did not find evidences of active star formation. Accordingly, IC~2944 and IC~2948 would represent different parts of the same, large group of clouds surrounding a cluster of OB stars \citep{Reipurth_2008}.
 \item More recently, \citet{Naze_et_al_2013} studied for the first time this region using XMM-Newton data and claimed for the existence of a tight cluster.
\end{itemize}

In this paper we perform a detailed wide-field study of the region based on $UBVIJHK$ data with other complementary spectroscopic and kinematic information. We use this material to derive updated estimates of the fundamental parameters of the various stellar groups. We also analyze this field in connection with other fields located on the fourth Galactic quadrant to derive information on the spiral structure in this portion of the disk.

The layout of the paper is as follows. In Sect.~\ref{sec:data} we describe the  data, and the reduction/calibration procedures. In Sect.~\ref{sec:analysis} we present our analysis of the data together with several discussions. Finally, in Sect.~\ref{sec:conclusions} we summarize our conclusions.

\section{Data} \label{sec:data}

This study makes use of the following material:

\begin{itemize}
\item Images, in $VI_C$ bands, obtained with the Wide Field Imager (WFI\footnote{http://www.eso.org/sci/facilities/lasilla/instruments/wfi.html}) mounted at the Cassegrain focus of the MPG/ESO 2.2m Telescope at La Silla Observatory (Chile).
\item Photometric data from the following surveys: APASS\footnote{http://www.aavso.org/apass} \citep{Henden_et_al_2010}, 2MASS\footnote{http://irsa.ipac.caltech.edu/Missions/2mass.html} \citep{Skrutskie_et_al_2006} and XHIP \citep{Anderson_Francis_2012}.
\item Spectroscopic classification for the brightest stars from the $SIMBAD$ and WEBDA databases, and complemented with information from \citet{Kharchenko_Roeser_2009}, \citet{Sana_et_al_2011} and \citet{Sota_et_al_2013}.
\end{itemize}

\subsection{Images}

The WFI camera is a 4 $\times$ 2 mosaic of 2K $\times$ 4K CCD detectors. The scale is $0\farcs238/pix$, and it  covers a field of view (FOV) of $34\farcm0 \times 33\farcm0$ but, due to the narrow inter-chips gaps, the filling factor of each image is $\sim$ 95.9\%. Images were acquired in 2006 during the nights of March 8$^{\rm th}$, April 25$^{\rm th}$ and May 28$^{\rm th}$ and the typical full width half maximum (FWHM) was in the range $0\farcs7 - 1\farcs4$, whereas air-masses values were 1.2 - 1.5. Details of the observations are given in Table~\ref{tab:frames}.

All frames were pre-processed in the standard way using the IRAF\footnote{IRAF is distributed by NOAO, which is operated by AURA under cooperative agreement with the NSF.}  task {\it ESOWFI/MSCRED}. That is, instrumental effects were corrected with calibration images (bias and sky-flats taken during the same observing runs).

\subsection{Astrometry}

World Coordinate System (WCS) header information of each frame was obtained using IRAF {\it MSCZERO} and {\it MSCCMATCH} tasks and UCAC4 data \citep{Zacharias_et_al_2013}. This allowed us to obtain a reliable astrometric calibration. In order to go deep with our photometry, all the long exposures for each band were combined using IRAF {\it MSCIMAGE} task. This procedure helps to both removing cosmic rays and improving the signal-to-noise ratio for the faintest stars.

\subsection{Photometry} \label{sec:photometry}

Instrumental magnitudes were extracted using IRAF {\it DAOPHOT} and {\it PHOTCAL} packages, and employing the point spread function (PSF) method \citep{Stetson_1987}. Since the FOV is large, a quadratic spatially variable PSF was adopted and its calibration on each image was done using several isolated, spatially well distributed, bright stars ($\sim$ 20) across the field. The PSF photometry was finally aperture-corrected for each filter and exposure time. Aperture corrections were computed performing aperture photometry of the same stars used as PSF models. Finally, all resulting tables from different filters and exposures were combined using {\it DAOMASTER} \citep{Stetson_1992} taking as reference the first night of observation (March 8$^{\rm th}$ 2006).

Data for stars in common with our observations (about 300) were selected from the APASS catalog and their Sloan $gri$ bands, and transformed to the $VI_C$ system using the equations provided by \citet{Jester_et_al_2005}. Then, we used the following transformation equations to obtain the corresponding $VI_C$ magnitudes from our instrumental ($vi$) ones:\\

\begin{center}
\begin{tabular}{llc}
$v = V + v_1 + v_2 (V-I_C)$   & ($r.m.s. = 0.05$) \\
$i = I_C + i_1 + i_2 (V-I_C)$ & ($r.m.s. = 0.07$) \\
\end{tabular}
\end{center}

\noindent with calibration coefficients presented in Table~\ref{tab:frames}.

\begin{table}
\fontsize{10} {14pt}\selectfont
\caption{Journal of observations of the scientific frames together
  with used calibration coefficients}
\centering
\begin{tabular}{llcc}
\hline
Date       & Frames & \multicolumn{2}{c}{Exposure times [sec]$\times N$} \\
\multicolumn{2}{c}{} & $V$   & $I_C$ \\
\hline
08/03/2006 & short  & -     & ~10x9~ \\
           & long   & 675x5 & 675x5~ \\
\hline
25/04/2006 & short  & ~10x9 & -      \\
           & long   & 675x5 & -      \\
\hline
28/05/2006 & short  & -     & -      \\
           & long   & -     & 675x10 \\
\hline
\hline
\multicolumn{4}{l}{Calibration coefficients (08/03/2006)} \\
\hline
\multicolumn {2}{l}{$v_1 = +1.044 \pm 0.006$} & \multicolumn {2}{l}{$i_1 = +1.951 \pm 0.009$} \\
\multicolumn {2}{l}{$v_2 = +0.070 \pm 0.006$} & \multicolumn {2}{l}{$i_2 = -0.007 \pm 0.010$} \\
\hline
\label{tab:frames}
\end{tabular}
\end{table}

\begin{table*}
\fontsize{10} {14pt}\selectfont
\caption{Completeness Factors (CFs) of optical data}
\centering
\begin{tabular}{ccccccccccc}
\hline
$V$       & 14-15 & 15-16 & 16-17 & 17-18 & 18-19 & 19-20 & 20-21 & 21-22 & 22-23 & 23-24 \\
\hline
$CF [\%]$ &   100 &   100 &   100 &   100 &    98 &    95 &    92 &    81 &    76 &    23 \\
\hline
\label{tab:cf}
\end{tabular}
\end{table*}

\begin{figure*}
\centering
\includegraphics[width=14cm]{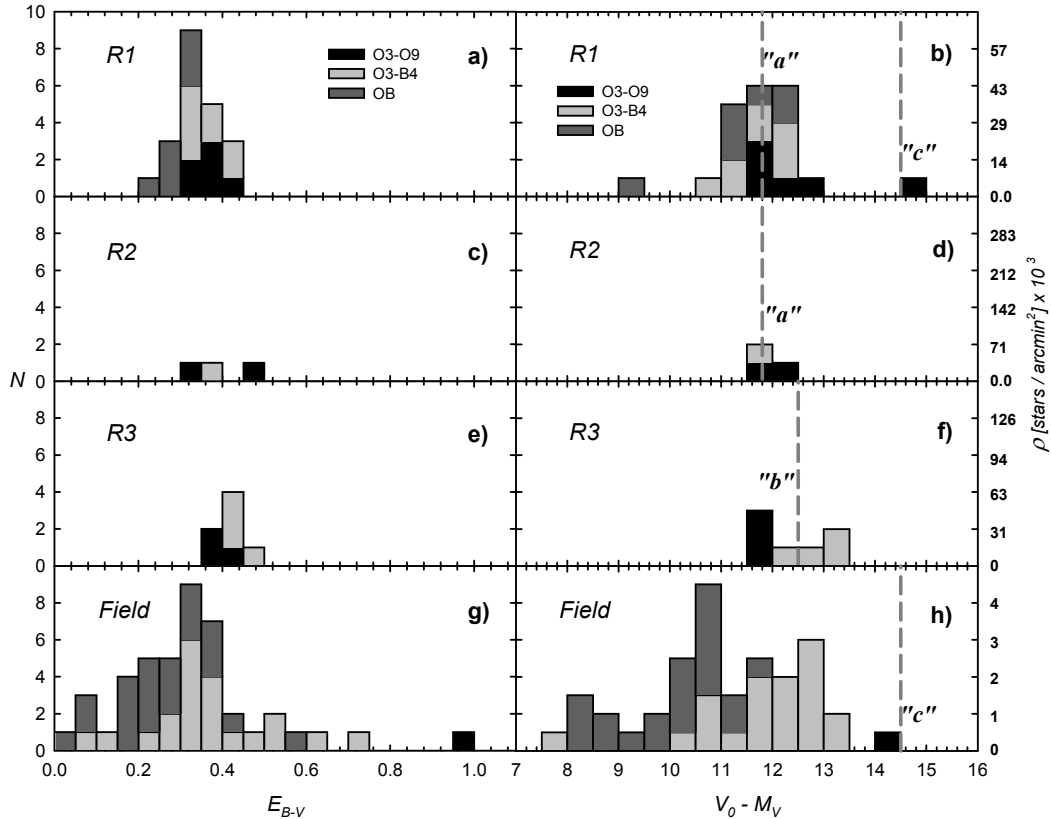}
\caption {Distributions of the $E_{B-V}$ and spectrophotometric $V_0-M_V$ of OB stars with known ST and located inside the selected regions (see Fig.~\ref{fig:dss}). Dashed lines in right panels indicate adopted $V_0-M_V$ values for stellar groups/populations placed in each region. For a better understanding, the highest bar of panel a) means: two O3-O9 stars, six O3-B4 stars and nine OB stars.}
\label{fig:spec}
\end{figure*}

In order to obtain a reliable initial mass function (IMF, Sect.~\ref{sec:IMF}), we estimated the completeness of our optical data. The procedure was already used in our previous works \citep[see e.g.][]{Baume_et_al_2003}. Basically, we created several artificial images by adding stars in random positions onto the original images using {\it ADDSTAR} routine of {\it DAOPHOT}. The added stars were distributed in luminosity as the real sample. In order to avoid overcrowding, in each experiment we added the equivalent to only about 15\% of the original amount of stars. Since $V$ band images are shallower than those in the $I$ band, we have adopted for he latter the completeness factor ($CF$) estimated for $V$. The $CF$ is defined then as the ratio between the number of artificial stars recovered and the number of artificial stars added,  and our results are listed in Table \ref{tab:cf},

\subsection{Kinematic data} \label{sec:kin}

Radial velocity information for stars in the zone were extracted from different bibliographic sources and databases. In detail, we obtained heliocentric radial velocities for 50 stars:
3 stars from $SIMBAD$ database,
19 from \citet{Kharchenko_et_al_2007} and \citet{Kharchenko_Roeser_2009},
20 from \citet{Huang_Gies_2006},
4 from \citet{Thackeray_Wesselink_1965},
1 from \citet{Conti_et_al._1977},
1 from \citet{Gies_et_al_2002} and
2 stars from \citet{Sana_et_al_2011}.
Regarding proper motions, all data were extracted from the UCAC4 catalog.

\subsection{Mid-IR and X ray data}

We cross-correlated our photometric data with mid-IR and X-ray data with the purpose of detecting pre-main sequence (PMS) stars (see also Sect~\ref{sec:lowms}).

As for the mid-R data, we use The Midcourse Space Experiment (MSX) data \citep{Egan_et_al_2003}. MSX mapped the galactic plane and other regions missed or identified as particularly interest by the Infrared Astronomical Satellite (IRAS) at wavelengths of $4.29 \mu m$, $4.35 \mu m$, $8.28 \mu m$, $12.13 \mu m$, $14.65 \mu m$, and $21.3 \mu m$.

The \citet{Naze_et_al_2013} catalog was used for the X-ray data. These data were obtained using the X-ray Multi-Mirror Mission Newton (XMM-Newton) information. This data resulted by binning its three energy bands: $soft=S=0.3-1.0~keV$, $medium=M=1.0-2.0~keV$, and $hard=H=2.0-10.0~keV$.

These data were cross-correlated with our optical data using $ALADIN$\footnote{http://aladin.u-strasbg.fr/} tool. To this aim, we use a $3\farcs0$ search radius for the XMM data, and a $10\farcs0$ radius for the MSX data --see \citet{Lumsden_et_al_2002} and \citet{Naze_et_al_2013}. We used the {\it all matches} option in $ALADIN$ since this option gave us all the counterparts within the searching  radius. Following these criteria, we found 139 counterparts in X-ray and 73 in mid-IR.

\subsection{Final catalog}

We used the STILTS\footnote{http://www.star.bris.ac.uk/~mbt/stilts/} tool to manipulate tables and we cross-correlated our $VI_C$ data with the other public $UBVIJHK$ photometry and the available spectral classification. In this process, we adopted ESO/WFI $VI_C$ data only for objects with $V > 13$. We obtained then a catalog with astrometric/photometric information of about 130,000 objects covering approximately a FOV of $44\farcm6 \times 48\farcm7$ around  IC~2944/2948 (see Fig.~\ref{fig:dss}). A more reliable analysis of the behavior of the Stellar Energy Distributions (SEDs) could be carried out with such a catalog (at least for brightest stars), thus preventing possible degeneracies in the photometric diagrams. The full catalog is presented in Table 3 which is  available in electronic form at Centre de Donn\'ees astronomiques de Strasbourg (CDS) web site.

\begin{figure*}
\begin{center}
\includegraphics[width=16cm]{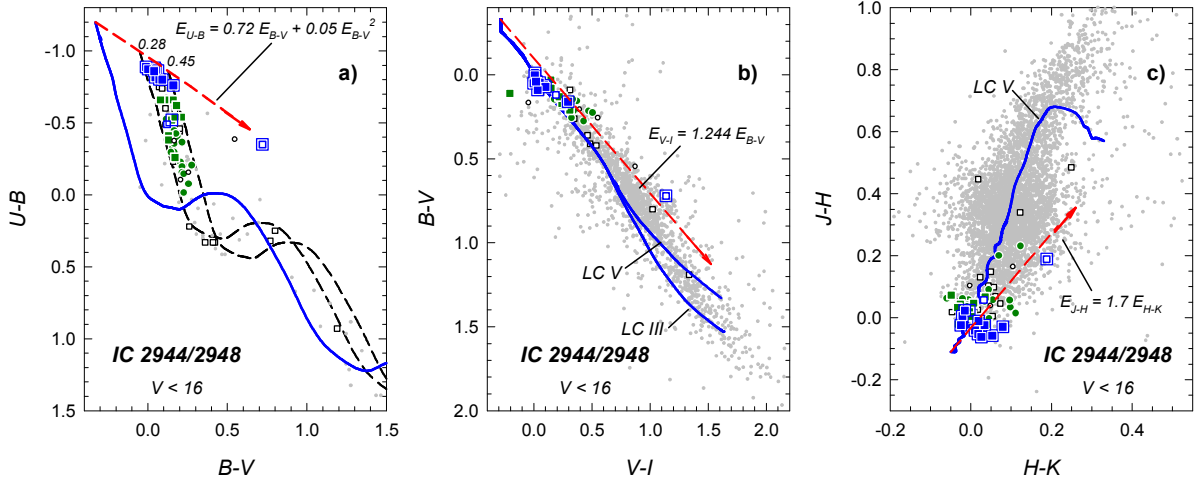}
\caption{TCDs for the brightest stars ($V<16$) in our sample. Symbols are as  in Fig.~\ref{fig:dss}b. The solid (blue) curve in plot a) is the \citet{Schmidt-Kaler_1982} ZAMS, while dashed (black) curves are the same ZAMS, but shifted along the reddening line (red) by the color excesses indicated above them.  Solid (blue) curves in plot b) are intrinsic colors for luminosity class V and III from \citet{Cousins_1978a,Cousins_1978b}. Solid (blue) curve in plot c) are intrinsic colors for luminosity class V from \citet{Koornneef_1983}. Dashed (red) arrows indicate the normal reddening path ($R_V = 3.1$).}
\label{fig:tcds}
\end{center}
\end{figure*}

\begin{figure*}
\begin{center}
\includegraphics[width=16cm]{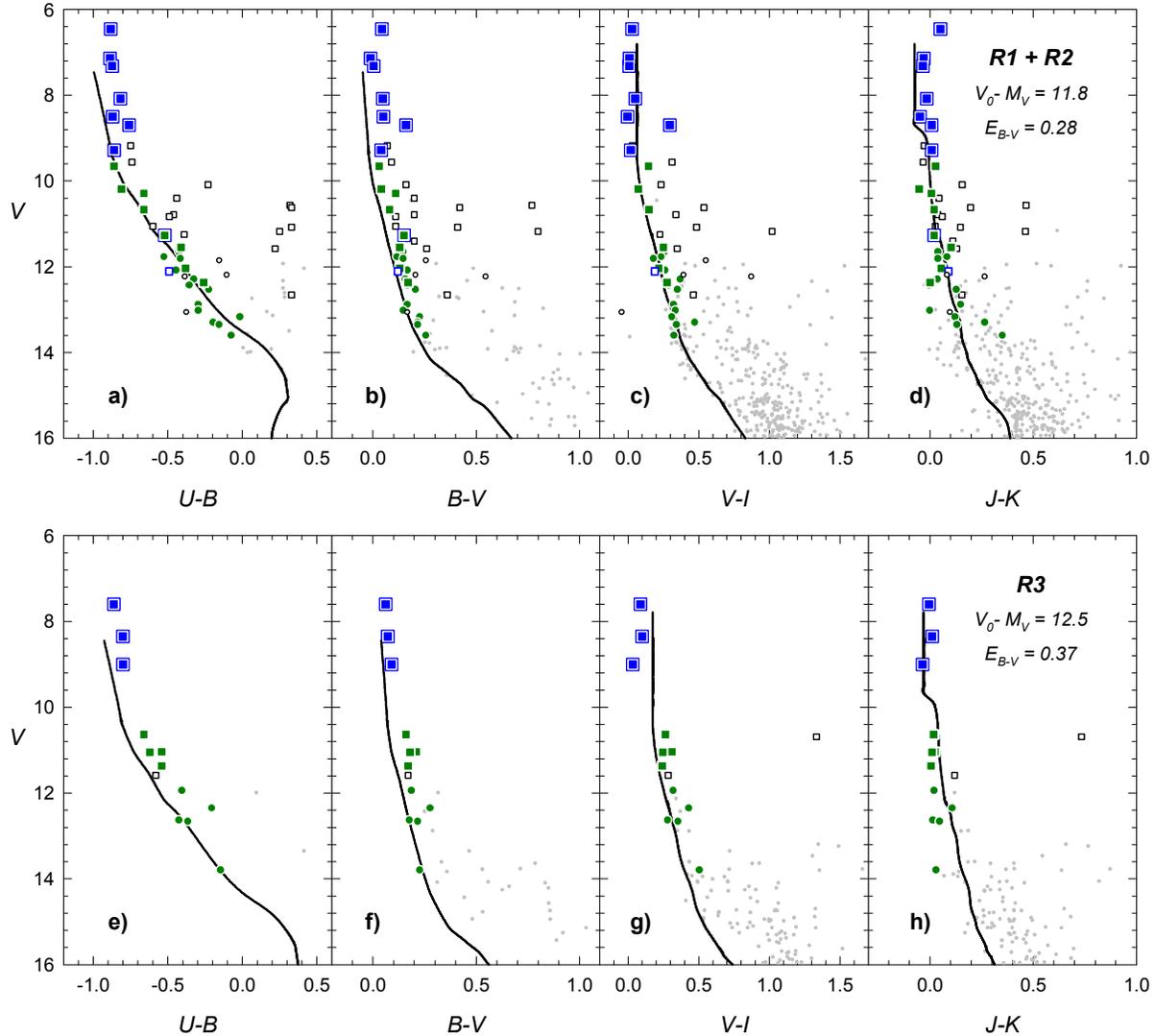}
\caption{CMDs of stars placed in the three selected regions of the studied zone (see Fig.~\ref{fig:dss}b). Symbols have the same meaning as in that figure. Solid (black) curves are the \citet{Schmidt-Kaler_1982} ZAMS (for $UBV$), the \citet{Cousins_1978a,Cousins_1978b} MS (for $VI$) and the \citet{Koornneef_1983} MS (for $JHK$) corrected by the respective adopted color excesses and apparent distance moduli for each stellar group located in each region (see Table~\ref{tab:param}).}
\label{fig:cmds}
\end{center}
\end{figure*}

\begin{figure*}
\centering
\includegraphics[width=15cm]{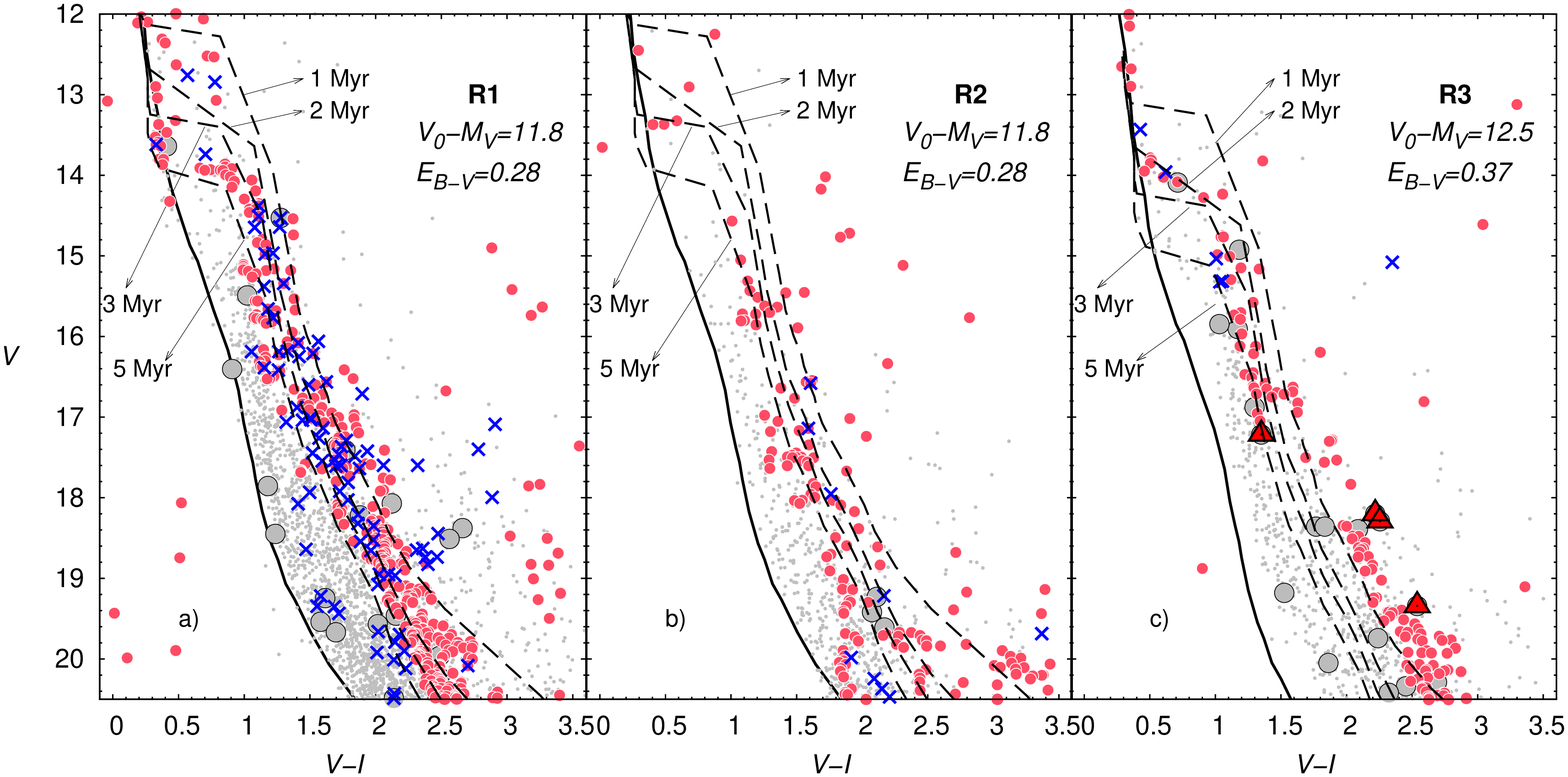}
\caption{Detailed CMDs of stars located in the three selected regions (grey dots). Solid curves are the \citet{Schmidt-Kaler_1982} and \citet{Cousins_1978a,Cousins_1978b} empirical MS; dashed curves are \citet{Siess_et_al_2000} isochones for $z = 0.02$. All the reference curves are corrected by the adopted color excesses values and apparent distance moduli (see Table~\ref{tab:param}). Red circles are stars resulting from statistical subtraction between each selected region and a comparison region. Blue 'X' are optical sources correlated with X ray sources. Big circles are optical sources correlated with MSX ray ones and red triangles are considered YSOs.}
 \label{fig:lms}
\end{figure*}

\section{Analysis and discussion} \label{sec:analysis}

\subsection{Selected regions} \label{sec:zones}

We started our study by identifying three regions where a visual over-density of early-type stars could be detected. These regions are shown as dotted circles in Fig.~\ref{fig:dss} and they are defined as follows:

\begin{itemize}
 \item {\it Region 1 (R1)}: A circle centered at $\Delta\alpha =   0\farcs0,
       \Delta\delta =    0\farcs0$ and radius $400\farcs0$
 \item {\it Region 2 (R2)}: A circle centered at $\Delta\alpha = -10\farcs0,
       \Delta\delta = +660\farcs0$ and radius $180\farcs0$
 \item {\it Region 3 (R3)}: A circle centered at $\Delta\alpha = 560\farcs0,
       \Delta\delta = -390\farcs0$ and radius $270\farcs0$
\end{itemize}

\noindent where we have used a coordinate system centered on HD~101205 star
($\Delta\alpha = (\alpha - \alpha_{HD101205})~cos(\delta_{HD101205})$;
$\Delta\delta = \delta - \delta_{HD101205}$).

\subsection{Spectrophotometric study} \label{sec:spec}

\subsubsection{Upper main sequence} \label{sec:upms}

Our compilation of spectroscopic data allowed us to estimate the distance and color excess of 162 stars located in our FOV. We applied then the traditional method \citep[e.g.][]{Corti_et_al_2012} based on the absolute magnitude $M_V$ and intrinsic colors calibration provided by \citet{Martins_Plez_2006} for O-type stars and by \citet{Schmidt-Kaler_1982} for B and later ones. We adopted an $M_V$ uncertainty of 0.5 magnitude \citep{Walborn_1972} and a value $R_V = A_V/E_{B-V} = 3.1$ (see in advance Fig.~\ref{fig:tcds}b). A linear interpolation was performed in case of missing calibration spectral type. In order to reduce the uncertainty, for binary systems, where both stars were classified (SB2), we obtained distance and color excesses by computing first the individual component magnitudes/colors and then we applied the above method. For binary systems with only one classification (SB1) we assumed our results have an additional minor uncertainty.

Computed distances and color excesses for known OB stars led us to the distributions presented in Fig.~\ref{fig:spec}, where we noticed the following features:

\begin{itemize}
\item Stellar densities (see scale on right axis) indicates a high concentration of young stars located in the selected regions in relation with the field values, confirming the criteria used in their choice.
\item Regarding the color excesses, the three regions are suffering some differential reddening, however it was possible to distinguish a concentration of values on each one. In particular, $R1$ and $R2$ this concentration is around $E_{B-V} = 0.33$ and in $R3$ it is around $E_{B-V} = 0.43$. On the other hand field stars reveals a wide spread of values with one O-type star with a strong reddening ($E_{B-V} \sim 1$).
\item Distance distributions reveal a clear peak in $R1$ and $R2$ at $V_0-M_V = 11.8$ and this value could be adopted also as a representative value for a possible stellar group ("a") placed in these regions. In $R3$ appears $V_0-M_V = 12.5$ as a better representative value for another possible stellar group ("b"). Additionally, there is one B-type star at about $V_0-M_V = 9.25$ located in $R1$, and there are two O-type star at about $V_0-M_V = 14$, one of them in $R1$ and the other in the field. These later two stars could be part of another group of young stars ("c"). There is also a set of B stars covering an important range in distance, from a few hundred pc to at least 1.5-2~kpc with an average stellar density $\sim 60\%$ higher than the average in the solar  neighbor-hood \citep{Allen_1973}.
\end{itemize}

\noindent
All these facts are reflected in the Two Color (TCDs) and Color Magnitude Diagrams (CMDs) shown in Figs.~\ref{fig:tcds} and \ref{fig:cmds}, respectively. In particular, Fig.~\ref{fig:tcds}a reveals the already indicated differential reddening with excess values ranging from 0.28 to 0.45, and Figs.~\ref{fig:tcds}b and \ref{fig:tcds}c indicate no evident deviations from the normal reddening law ($R = A_V/E_{B-V} = 3.1$) in the zone, in agreement with previous polarimetric studies \citep[see][]{Vega_et_al_1994}.

\addtocounter {table} {1}
\begin{table*}
\fontsize{10} {14pt}\selectfont
\caption{Parameters of main stellar groups}
\centering
\begin{tabular}{lccc}
\hline
\multicolumn{1}{c}{Parameter}       & $"a"$            & $"b"$            & $"c"$       \\
\hline
$V_0-M_V$                           & 11.8 $\pm$ 0.2   & 12.5 $\pm$ 0.2   & $\sim 14$   \\
$E_{B-V}$                           & 0.28-0.40        & 0.37-0.45        & $\sim 1~?$  \\
\hline
$Nuclear~Age$ [Myr]                 & $\sim 3$         & $\sim 3$         & $\sim 10~?$ \\
$Contraction~Age$ [Myr]             & $\sim 3-5$       & $\sim 2-3$       & -           \\
\hline
$IMF~slope~(\Gamma)$     (case a)   & $-1.00 \pm 0.35$ & $-1.02 \pm 0.37$ & -           \\
~~~~~~~~~~~~~~~~~~~~~~~~ (case b)   & $-0.94 \pm 0.27$ & -                & -           \\
$Total~mass$ [M$_{\odot}$] (case a) & $\sim1100\pm200$ & $\sim470\pm100$  & -           \\
~~~~~~~~~~~~~~~~~~~~~~~~~~ (case b) & $\sim1200\pm200$ & -                & -           \\
\hline
\hline
\multicolumn{4}{l}{Kinematic values from O member stars} \\
\hline
$\mu_{\alpha}\cos(\delta)$ [mas/yr]& $-5.1\pm1.2~(7)$~~~~  & $-6.1\pm1.1~(2)^*$ & $\sim -0.2~?~(2)$ \\
$\mu_{\delta}$ [mas/yr]            & $+0.2\pm1.2~(7)$~~~~  & $-0.7\pm0.8~(2)^*$ & $\sim -5.6~?~(2)$ \\
$RV_{LSR}$ [km/s]                  & $+5.9\pm2.8~(5)^{**}$ & $+4.4\pm0.5~(2)$~~ & $\sim-10.7~?~(1)$ \\
\hline
\label{tab:param}
\end{tabular}
\begin{center}
\vspace{-0.6cm}
\begin{minipage}[c]{11cm}
\fontsize{9} {11pt}\selectfont
{\bf Notes:} \\
- Number of stars is indicated in parenthesis \\
- * Star HD 101436 excluded \\
- ** Stars HD 101205 and HD 101223 excluded \\
- IMF case a: HD 101205 star was not considered as member \\
- IMF case b: HD 101205 star was considered as member \\
- ?: very uncertain values \\
\end{minipage}
\end{center}
\end{table*}

\begin{figure*}
\centering
\includegraphics[width=14cm]{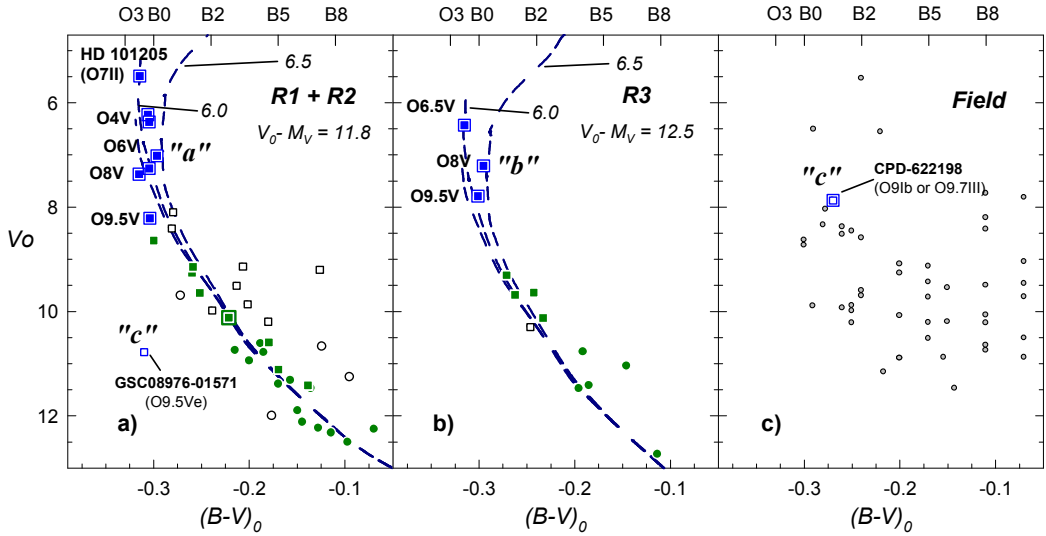}
\caption{Corrected CMDs of stars presented in Fig.~\ref{fig:cmds}. Symbols are as in Fig.~\ref{fig:dss}b. Dashed (blue) curves are \citet{Marigo_et_al_2008} isochrones for $z = 0.02$ (numbers indicates log(age[yr])) . All the reference curves are corrected by the respective adopted distance modulus (see Table~\ref{tab:param}). Upper axis indicate ST for MS stars. O stars representative of each stellar group/population are indicated with letters $"a"$, $"b"$ and $"c"$.}
\label{fig:cmds0}
\end{figure*}

\subsubsection{Lower main sequence} \label{sec:lowms}

To study the lower main sequence (MS) ($V > 14$) we first attempt to quantify field star contamination. To this purpose, we developed a code that performed a statistical decontamination of the field stars over the CMDs \citep[see][]{Gallart_et_al_2003}. Briefly, the procedure consists in a star-by-star position comparison onto the CMDs of each region ($R1$, $R2$ and $R3$) and a CMD of a representative field located near them and covering the same sky area. Then, we subtract field stars with similar position (color and magnitude)  from the  CMDs to obtain a decontaminated diagram. For this task we used our $V$ vs. $V-I$ diagrams since they are the deepest ones. In Fig.~\ref{fig:lms} we present the resulting CMDs, where we could identify a highly probable PMS population (small red circles). To strengthen our results, we use X-ray data as an additional membership criteria. PMS stars are routinely identified using X-ray emission and this kind of observations should reveal almost all ($\sim 90\%$) the PMS population up to the corresponding detection limit \citep{Feigelson_Getman_2005}. Different panels in Fig.~\ref{fig:lms} (x symbols) reveal their spatial distribution over the three selected regions. We also cross-correlated our data with IR sources present inside these three regions (big grey circles in Fig.~\ref{fig:lms}). In particular, we mark those ones (red triangles) with infrared behavior expected for a young stellar objects (YSOs), say with an infrared rising energy distribution -- $F_{21}/F_{14} > 1$ and $F_{14}/F_8 > 1$, \citep{Lumsden_et_al_2002}.

We could remark the following:
a) Most  optical sources that could be associated with an X-ray counterpart follow closely the color and magnitude distribution of the stars we selected as probable PMS objects after the field star decontamination.
b) The observed width in both set of data could be caused by an age spread, binary stars, accretion stellar discs -- \citet{Preibisch_Zinnecker_1999}; \citet{Kenyon_Hartmann_1990}-- or/and differential reddening.
c) The amount of X-ray sources is significantly lower than those revealed in the decontamination process; and
d) We found only a few probable YSOs (Fig.~\ref{fig:lms}c).

\subsection{Stellar groups and selected stars} \label{sec:groups}

The analysis in Sect.~\ref{sec:upms} allowed us to better understand this particular galactic direction. Previous work by other authors (see Sect.~\ref{sec:intro}) claimed the presence of a spread of early stars along the line of sight or the presence of several stellar groups. Our analysis indicated that we could recognize a spread of early (B-type) star, two main stellar groups and a young background stellar population. Each group exhibit  an important dispersion in distance/excesses, that  one can explain with spectroscopic errors, and other well known effects such as  multiplicity, evolution, fast rotation and/or differential reddening. The main stellar groups/populations can be described as follows:

\begin{itemize}
 \item $"a"$: This stellar group is located at about 2.3~kpc from the Sun, suffering an $E_{B-V}$ ranging from 0.28 to 0.40, and mainly located in $R1$. However, the brightest stars of $R2$ revealed similar properties and they were also considered part of the same stellar group.
 \item $"b"$: This stellar group is at about 3.2~kpc from the Sun, suffering an $E_{B-V}$ ranging from 0.37 to 0.45, and mainly located in $R3$.
 \item $"c"$: This is an apparent spread of young background population at about 8~kpc from the Sun, represented in our data by at least two O-type stars.
\end{itemize}

Under these assumptions and adopting a normal reddening law, we simultaneously fit a zero main sequence (ZAMS) or a MS in all the CMDs (see Fig.~\ref{fig:cmds}) at respective main groups distances revealed in Fig.~\ref{fig:spec}. These relationship closely follow both the brightest stars and the fainter B-type ones.

In Fig.~\ref{fig:cmds0} we reddening-present the corrected CMDs for brightest ($V < 14$) stars located in $R1+R2$, $R3$ and the field around them in different panels.We use  only stars with available spectral classification, or for which it  was possible to apply  the classic de-reddening method \citep[see][]{Baume_et_al_2003} and the relation $E_{U-B}/E_{B-V} = 0.72 + 0.05~E_{B-V}$ (see Fig.~\ref{fig:ccds}a). We noticed that in Fig.~\ref{fig:cmds0} stars located in $R1+R2$ and $R3$ show consistent magnitude and spectral classification sequences and it seems conceivable to associate most OB stars in each panel with the stellar groups $"a"$ and $"b"$ mentioned above.

A few stars deserve additional comments (see also Fig.\ref{fig:cmds0}):
\begin{itemize}
\item HD~101205: It is the brightest star in the selected regions and is classified at $SIMBAD$ as a O8~V but as a O7~II by \citet{Sota_et_al_2013}. In both cases this star do not follow the above indicated sequence and under the former classification it yields an spectroscopic $V_0-M_V = 10.2$. However, adopting the later classification and considering it is a quadruple system with at least 3 OB components \citep{Sana_et_al_2011} it is possible to explain its particular location on Fig.\ref{fig:cmds0} and could be considered as a group $"a"$ member.
\item CPD-62${\degr}$2198: This star is placed among other well located MS O-type stars but it is located in the field, outside the selected regions $R1$, $R2$ and $R3$. This star is classified as a supergiant (O9~Ib; WEBDA) or as a giant \citep[O9.7~III;][]{Sana_et_al_2011} revealing it could have a spectroscopic distance moduli of 14.2 or 13.0, respectively. In any case this star would be behind the main stellar groups and it would  be part of the called population $"c"$.
\item HD~101333: This star is also located along a MS (see Fig.\ref{fig:cmds0}), however it is classified as a giant (O9.5~III; $SIMBAD$) or a dwarf \citep[O9.5~V;][]{Sana_et_al_2011}. In the former classification it would share the place with CPD-62${\degr}$2198, but, if a giant,  it could still be considered part of the $"b"$ group.
\item HD~308804: This star has also different spectral classifications. It is a O9.5~Vn in $SIMBAD$ and a B3~V by \citet{Sana_et_al_2011}. As an O-type star, it would have a similar distance as CPD-62${\degr}$2198. However, by adopting the later spectral type, it could be considered member of the  $"a"$ group.
\item GSC08976-01571: This star is classified as a O9.5~Ve (WEBDA). It is placed in $R1$ but it is a too faint MS star to belong to the main stellar groups. With a spectroscopic distance $V_0-M_V = 14.8$  it  be a member of the $"c"$ group.
\end{itemize}

The resulting parameters of the main stellar groups and background stellar population are summarized in Table~\ref{tab:param}.

\subsection{The Galactic spiral structure} \label{sec:galactic}

Previous investigations --\citet{Carraro_Costa_2009}, \citet{Baume_et_al_2009, Baume_et_al_2011} and references therein-- presented their analysis of optical observations along directions close of the field studied in this paper. In all cases, several early-type star groups (usually three) were detected along the line of sight, and they were considered as belonging to spiral features located at increasing distance from the Sun.

In this work we found several populations of young stars as well, thus lending further support to our earlier results. In this study we  found: a) a population represented by several B-type stars scattered form a few hundred of parsecs to at least 1.5-2~kpc of the Sun; b) a second population identified by the young stellar groups $"a"$ and $"b"$ with several O stars situated at 2.3 and 3.2~kpc respectively; and c) a third population which is represented by at least two O stars (one in $R1$ and other in the field) identified as population $"c"$ and located at about 8~kpc.

\begin{figure}
\centering
\includegraphics[width=8cm]{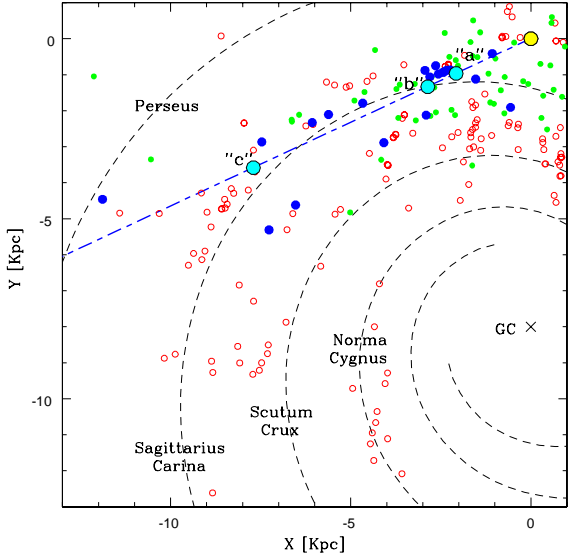}
\caption{Comparison of the studied Galactic direction (dashed blue line) to the \citet{Vallee_2008} model with four arms. The big cyan open circles indicate the studied "a" and "b" stellar groups and the background young stellar population "c" (see Sect.~\ref{sec:groups}). We assumed a distance to the Galactic Centre of $R_0$ = 8.0~kpc. Other stellar populations are plotted for reference: (green) points are Cepheid \citep{Majaess_et_al_2009}; filled (blue) circles are open clusters and young groups from optical ($UBVI$) data (from several studies); and open small (red) circles mark the HII tracers \citep{Hou_et_al_2009}. The yellow circle marks the position of the Sun.}
\label{fig:arms}
\end{figure}

The results of this paper together with previous results are presented in Fig.~\ref{fig:arms} where the \citet{Vallee_2008} spiral arms' model is also shown as dotted curves. It reveals then a consistent picture between the described groups/populations and those already known. It seems necessary to study more directions close to this regions using optical and infrared observations of young objects to complete the picture of spiral structure in the fourth Galactic quadrant.

\begin{figure}
\centering
\includegraphics[width=8cm]{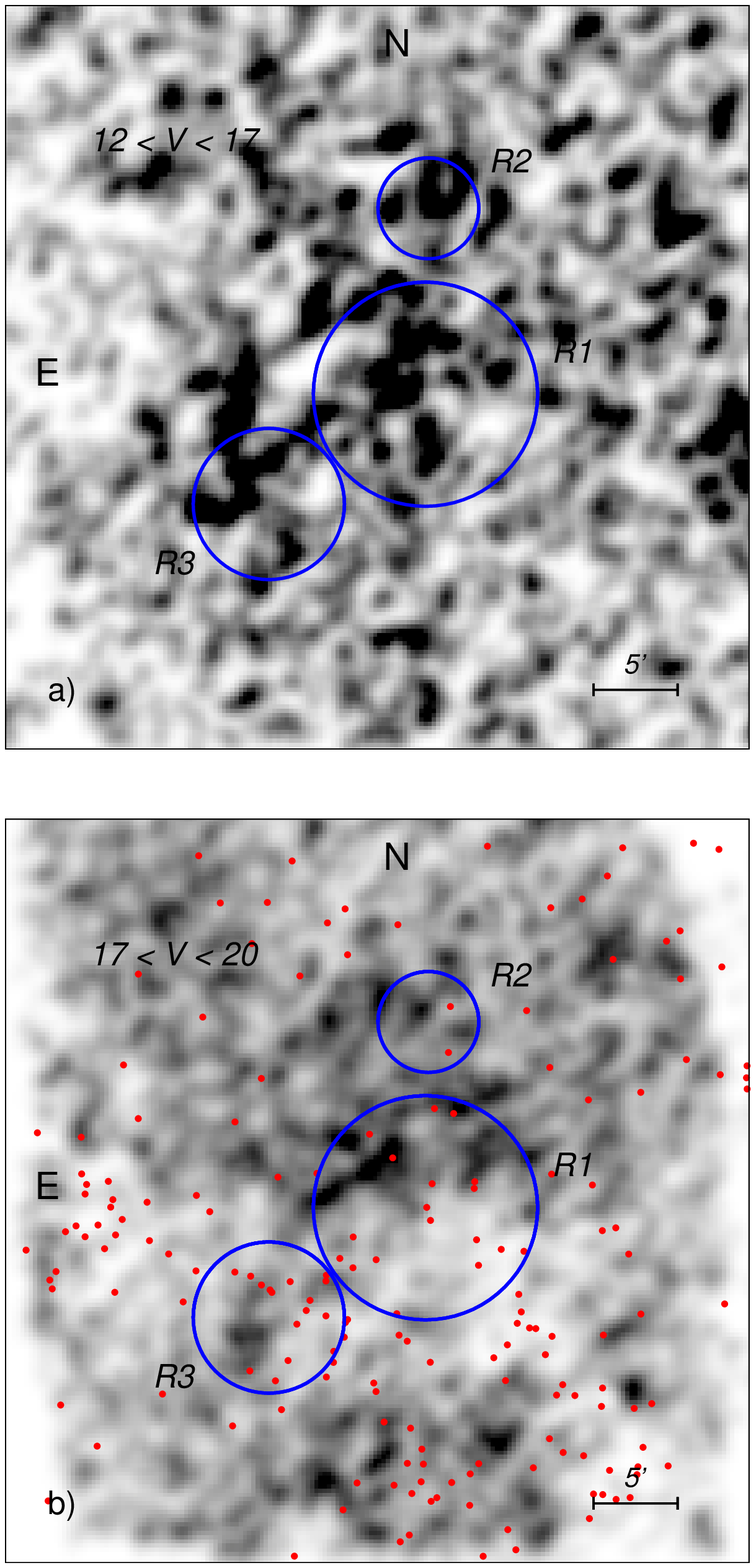}
\caption{Density maps for stars located between two PMS isochrones (see text). Both panel discriminate stars brighter and fainter than V = 17. Blue circles indicate the selected regions. Red dots are MSX sources \citep{Lumsden_et_al_2002} correlated with our optical data.}
\centering
\label{fig:density}
\end{figure}

\subsection{Dating the stellar groups}

In order to estimate the ages of the main stellar groups, we compared our data with different theoretical sets of isochrones over the CMDs and we also took into account the spectral stellar classifications (when available). We considered then two cases:

\begin{itemize}
\item For brightest (earlier) stars, we compared their CMD position (see Fig.~\ref{fig:cmds0}) with models computed for solar metallicity, mass loss and overshooting \citep{Marigo_et_al_2008} shifted by adopted distance values for groups "a" ($V_0-M_V = 11.8$) and "b" ($V_0-M_V = 12.5$). Intrinsic scatter is present in our data and this fact prevents a unique isochrone solution, however it is still possible to obtain an estimate of the age value for stellar groups $"a"$ and $"b"$, say $\sim$ 3~Myr. This value is consistent with the expected lifetime on the MS of the earlier stars of each group ($\sim$ O5). These values obtained for post MS evolution were identified as "nuclear ages". For population "c", we adopted the lifetime of a O9-B0 on MS ($\sim$ 10~Myr) as age estimate.
\item For the faintest (PMS population) stars, we superimposed a set of \citet{Siess_et_al_2000} PMS models on CMDs presented in Fig.~\ref{fig:lms} using the respectives $E_{B-V}$ and $V_0-M_V$ values. The transition from PMS to MS phase is also an extremely accurate age diagnostic. This phase is clearly seen in $R1$ and in $R3$, and to less extent  in $R2$ and indicates an age between 2 and 3~Myr. The adopted age values for each group are indicated in Table~\ref{tab:param} identified as "contraction ages".
\end{itemize}

Taking into account the uncertainties in the determinations of the ages and distances, it is possible to conclude that there is no significant difference between the more massive and less massive stars' ages.

\subsection{PMS population}

Regarding the PMS population, we noticed that most of X-ray sources are distributed uniformly inside $R1$ but tend to show un-even distributions between $R1$ and the regions $R2$ and $R3$, suggesting separated groups. On the other hand, identified YSOs are all placed in $R3$ (see Figs.~\ref{fig:dss}b and ~\ref{fig:lms}). This difference could be in part real and indicate a different evolutionary stage, but it is more conservatively  caused by a distance objects population placed in this region as was suggested by the presence of a recently discovered embedded clusters using data from the deep near infrared VVV \footnote{Vista Variables in the Via Lactea} survey \citep{Borissova_et_al_2011}.

In order to better understand the PMS population distribution in the region, we build the spatial density stellar maps selecting stars placed between the following limits: a PMS isochrone of 1~Myr (at $E_{B-V} = 0.28$ and $V_0-M_V = 11.8$) and a PMS isochrone of 5~Myr (at $E_{B-V} = 0.37$ and $V_0-M_V = 12.5$). In this process, we used first the triangles method \citep{Kippenhahn_1967} for the selection over the CMDs of all the observed stars, and then we choose a spatial bin size of $1\farcm0$ and the drizzle method \citep{Fruchter_Hook_2002} with a $15\farcs0$ step for the building of the density maps. Our results are presented in Fig.~\ref{fig:density} where we distinguish stars brighter and fainter than $V = 17$ ($\sim$ 1.6-2 M$_{\odot}$). We noticed that $V < 17$ stars (Fig.~\ref{fig:density}a) present over-densities well correlated with the previous selected regions according the MS stars ($V < 12$). On the other hand, $V > 17$ stars (Fig.~\ref{fig:density}b) only presented an appreciable over-density in the north part of $R1$. We also observed a close correlation between lower density regions in this plot and MSX sources -- mainly located at south and east of $R1$-$R3$. This correlation indicates that this map mainly reflects the differential absorption present across our FOV and not real stellar concentrations.

All these facts reinforce the idea that the PMS population in the studied field is nearly 2~Myr, mainly concentrated in $R1$ and could be revealed in our analysis up to $V = 17$ . Fainter (less massive) possible PMS population could not be separated from the field population using this method.

\subsection{The initial mass function of the main stellar group} \label{sec:IMF}

\begin{figure}
\centering
\includegraphics[width=8cm]{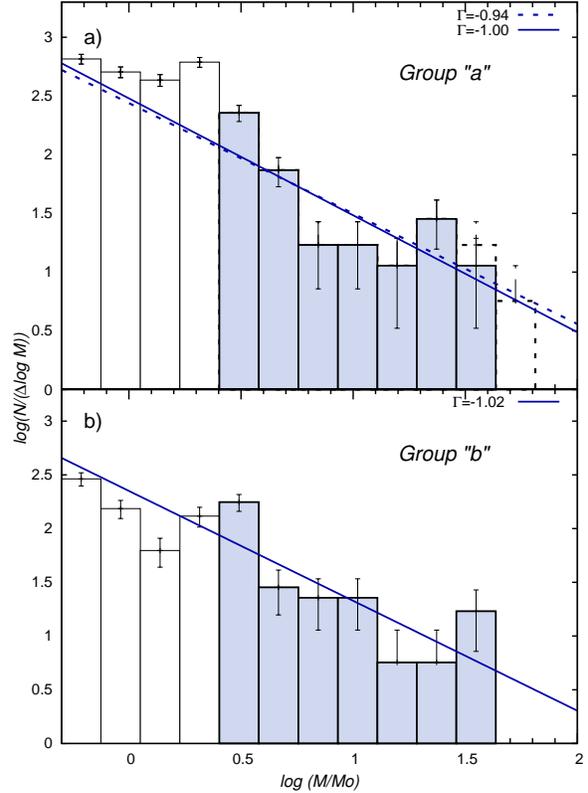}
\caption{IMFs obtained for the stellar groups "a" and "b" (see Sect.~\ref{sec:groups}). For group "a", both cases represent results were HD~101205 was and was not included. Grey bins (M $>$ 2~M$_{\odot}$) represent values considered to compute the power law fits. Error bars indicate Poisson uncertainty.}
 \label{fig:imf}
\end{figure}

The Initial Mass Function (IMF) can be approximated by a power law of form: \\

log ($N/\Delta~(log~m)$) = $\Gamma$ log $m$, \\

\noindent
where $N$ is the number of star per logarithmic mass bin $\Delta(log~m)$.

We computed  the IMFs for groups "a" and "b", using respectively stars located in regions $R1+R2$ and $R3$ (see Sect.~\ref{sec:groups}). To build the IMFs, we divided, on each region, the stars in two sets: brighter and fainter than $V = 12$. In the former case, we considered only stars adopted as members following our spectrophotometric analysis (see Sect.~\ref{sec:spec}) and individual masses were estimated via a linear interpolation the \citet{Marigo_et_al_2008} model for 3~Myr. For the latler case, we considered the stars resulting from the statistical decontamination (see Sect.~\ref{sec:lowms}) and we used the \citet{Siess_et_al_2000} stellar model for 3~Myr for deriving individual masses. We also considered the CFs (see Sect.~\ref{sec:photometry}) which, however,  are important only at very low mass values (M $<$ 0.5 M$_{\odot}$). Figure \ref{fig:imf} shows the resulting IMFs (histograms) and the least square fits to the data (lines). In all these fits we took into account only stars with masses larger than 2~M$_{\odot}$. The IMF slopes are reported in Fig.~\ref{fig:imf} and Table \ref{tab:param}. These values nicely compares (within the errors) with the \citet{Salpeter_1955} slope ($\Gamma = -1.35$), although they tend to be lower than this value.

We estimated the total mass of the stellar groups integrating over the mass range as follows: individual masses for all the stars with masses greater than 2 M$_{\odot}$ were summed up, while for lower masses we estimated the amount of star at each mass bin using the values given by the power law. The estimated total masses for each stellar group are presented in Table~\ref{tab:param}.

We repeated the analysis, but using for the faint group only stars with a XMM counterpart. In this case we obtained for each bin smaller numbers of sources ($\sim 15\% - 30\%$), as was expected from a simple visual inspection of Fig.~\ref{fig:lms}. This difference might derive from X-ray data incompleteness and crowding, as previously suggested  by \citet{Naze_et_al_2013}. A similar situation was reported by \citet{Prisinzano_2005} in their study of NGC~6530.

\subsection{Kinematic study}

We extracted heliocentric radial velocities from the WEBDA database for 25 stars (see Sect.~\ref{sec:kin}). Since they come from different sources, taken at different epochs and with different spectral dispersions, we averaged the values obtained from better spectral resolution observations. For binary systems, their barycentric radial velocities were computed analysing the best-fitting curves from \citet{Gies_et_al_2002} and \citet{Sana_et_al_2011}.

In order to compare the stellar radial velocities with a model of Galactic rotation, we corrected the heliocentric values to the Local Standard of Rest (LSR). We adopted 13.3 km/s for the solar motion relative to the LSR -- $(U,V,W)_{\odot} = (9.96, 5.25, 7.07)$ km/s -- according to \citet{Aumer_Binney_2009} and we used the stellar spectrophotometric distances computed in Sect.~\ref{sec:spec}.

\begin{figure}
\centering
\includegraphics[width=8cm]{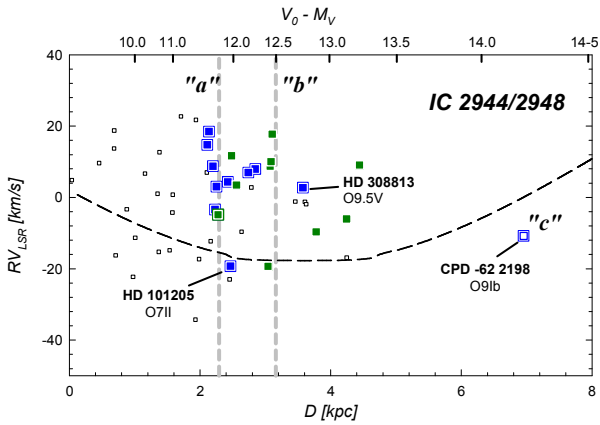}
\caption{AL fit of the galactic rotation model (dashed curve) applied to our Galaxy \citep{Fich_et_al_1989} and LSR radial velocities of stars placed in the studied zone (see Fig.~\ref{fig:dss}). Symbols are the same meaning as in Fig.~\ref{fig:dss}b. Vertical lines indicates the adopted 2.3 and 3.2~kpc distances ($V_0-M_V = 11.8$ and $V_0-M_V = 12.5$) for the stellar groups $"a"$ and $"b"$ (see Sect.~\ref{sec:groups})}
\label{fig:VrvsD}
\end{figure}

As for the Galactic rotation curve, we considered the Adjusted Linear Model (AL) and the Power Law Model \citep{Fich_et_al_1989}. Since, in our case, both models provide similar results, we eventually chose the former. Results are presented in Fig.~\ref{fig:VrvsD}, We computed then the mean LSR velocities of the groups "a" and "b"  considering only their O stars. In the case of the population "c", only CPD-62${\degr}$2198 star provided radial velocity data and it was adopted as a poor representative value for these stars (see Table~\ref{tab:param}). We then could conclude that stellar groups $"a"$ and $"b"$ (at about 2.3 and 3.2~kpc respectively) do not follow the Galactic rotation model and it is possible that population "c" could have a closer behavior. We noticed that the particular kinematic behavior in radial velocities of the studied stellar groups $"a"$ and $"b"$ could may have been mistakenly interpret an individual member as a run away star. In fact, it could be the case of star HD~101131, since this one was identified as a run away object by \citet{van_Buren_1995}. However, recently this star was only identified as an object with an associated extended source \citep{Peri_et_al_2012} with an excess 60 $\mu$ emission that could be produced by a bow-shock \citep{Noriega-Crespo_et_al_1997}.

\begin{figure}
\centering
\includegraphics[width=7cm]{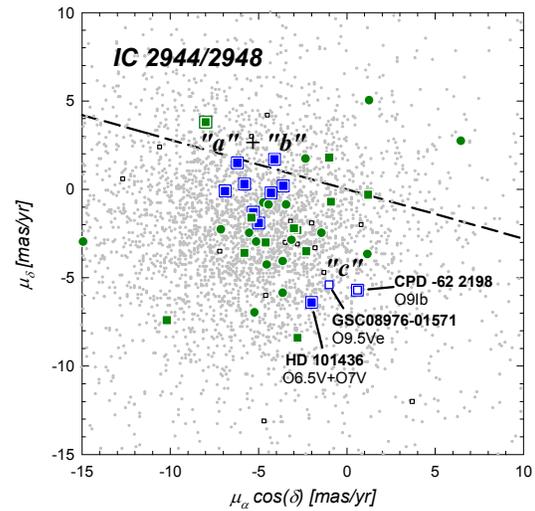}
\caption{VPD for the stars brigther than $V = 16$. Symbols are the same meaning as in Fig.~\ref{fig:dss}b. Dashed line indicates expected direction for movements parallel to the Galactic plane.}
\centering
\label{fig:mpr12.5}
\end{figure}

The cross-correlation of our photometric data against the UCAC4 catalog allows us to analyze also the proper motion of more than 7500 stars in the studied area (see Fig.~\ref{fig:dss}). Figure \ref{fig:mpr12.5} presents the Vector Point Diagram (VPD), where we marked previously identified stars (as in Fig.~\ref{fig:dss}b). Most of O stars considered stellar groups $"a"$ or $"b"$ members show quite a similar motion. On the other hand, O stars identifying population $"c"$ (hollow blue symbols) seems to show again a different kinematic behavior. We adopted then the averaged proper motions of O stars of each group as representative of each one and are presented in Table~\ref{tab:param}.

The kinematic information for brightest stars in the studied region was summarized in Table~\ref{tab:kin}.

\begin{table*}
\fontsize{10}{13pt}
\leavevmode
\caption{Kinematic information for studied OB stars}
\centering
\label{tab:kin}
\begin{tabular}{lrlcrlrlll}
\hline
 Name &
 \multicolumn{2}{c}{RV$_{LSR}$} &
 \multicolumn{1}{c}{N$^{(*)}$} &
 \multicolumn{2}{c}{$\mu_{\alpha}cos{\delta}$}&
 \multicolumn{2}{c}{$\mu_{\delta}$}&
 Binarity & RV Source \\
      &
 \multicolumn{2}{c}{[km s$^{-1}$]} & &
 \multicolumn{2}{c}{[mas yr$^{-1}$]} &
 \multicolumn{2}{c}{[mas yr$^{-1}$]} & & \\
\hline
HD101205            & -19 & (20) &    &  -6.2 & (1.0) &  1.5 & (1.0) & SB2$^{~b}$ & Kharchenko (2007) \\
HD101131            &   9 & (11) & 75 &  -5.8 & (1.0) &  0.3 & (1.0) & SB2$^{~a}$ & Gies et al. (2002) \\
HD101190            &   8 &  (9) & 20 &  -4.1 & (1.0) &  1.7 & (1.0) & SB2$^{~b}$ & Sana et al. (2011) \\
HD101436            &   4 & (10) & 13 &  -2.0 & (1.6) & -6.4 & (4.2) & SB2$^{~b}$ & Sana et al. (2011) \\
HD101298            &   7 &  (8) &  3 &  -3.6 & (1.2) &  0.2 & (1.2) & SB?$^{~b}$ & Trackeray \& Wesselink (1965) \\
HD101413            &  18 & (16) & 14 &  -6.9 & (1.3) & -0.1 & (1.4) & SB2$^{~b}$ & Conti et al. (1977) \\
HD101191            &   3 &  (7) &  3 &  -6.9 & (1.2) & -0.1 & (1.2) & SB$^{~b}$  & Trackeray \& Wesselink (1965) \\
HD101223            &  15 &  (7) &  3 &  -5.0 & (1.3) & -1.9 & (1.7) & SB?$^{~b}$ & Trackeray \& Wesselink (1965) \\
HD101333            &  -3 &  (5) &  3 &  -5.3 & (1.3) & -1.3 & (1.3) & SB?$^{~b}$ & Huang \& Gies (2006) \\
HD308813            &   3 & (22) &  3 &  -4.3 & (1.4) & -0.2 & (1.3) & SB$^{~b}$  & Huang \& Gies (2006) \\
HD308818            &   9 &  (7) &    &  -2.8 & (5.1) & -2.3 & (1.3) &            & Karchenko (2007) \\
CPD-62${\degr}$2153 &  18 &  (7) &    &  -2.3 & (1.1) & -3.5 & (1.3) &            & Karchenko (2007) \\
HD308826            &  12 &  (-) &    &  -3.0 & (2.3) & -2.2 & (2.0) &            & SIMBAD \\
HD308825            &  -6 &  (0) &  3 &  -4.6 & (1.3) & -3.0 & (3.1) &            & Huang \& Gies (2006) \\
HD308817            &  10 &  (9) &  3 & -10.2 & (1.4) & -7.4 & (1.6) &            & Huang \& Gies (2006) \\
CPD-62${\degr}$2198 & -11 &  (4) &  3 &   0.6 & (1.1) & -5.7 & (1.0) & SB?$^{~b}$ & Huang \& Gies (2006) \\
HD308831            & -19 & (10) &  3 &  -1.0 & (3.1) &  1.8 & (3.5) &            & Huang \& Gies (2006) \\
HD308833            & -10 &  (9) &  3 &   1.2 & (1.7) & -0.3 & (2.4) &            & Huang \& Gies (2006) \\
HD308804            &  -5 &  (3) &  3 &  -8.0 & (1.5) &  3.8 & (2.6) & SB?$^{~b}$ & Huang \& Gies (2006) \\
HD308832            &   9 &  (2) &  3 &  -5.8 & (1.8) & -3.6 & (1.7) &            & Huang \& Gies (2006) \\
GSC08976-00983      &   3 &  (7) &  3 &  -0.9 & (1.9) & -0.7 & (1.7) &            & Huang \& Gies (2006) \\
\hline
\end{tabular}
\begin{center}
\begin{minipage}[c]{14cm}
\fontsize{9} {11pt}\selectfont
{\bf Notes:} \\
- (*) = Number of averaged spectra \\
- Errors are presented in brackets \\
- Binarity source: $a$ = \citet{Gies_et_al_2002}; $b$ = \citet{Sana_et_al_2011} \\
- SB = Binary system, single spectra; SB2 = Binary system, double spectra;
  SB? = Dubious binary system (possible binary systems by
  \citet{Ardeberg_Maurice_1977} but probable single star by \citet{Sana_et_al_2011})
\end{minipage}
\end{center}
\end{table*}

\section{Conclusions} \label{sec:conclusions}

In this study we reported on a deep optical photometric study of the region surrounding IC~2944/2948 at ($l \sim 294\fdg8$; $b \sim -1\fdg6$). Optical photometry was complemented with archival data, which include astrometry, shallow optical, IR and mid-IR photometry, and spectroscopy. This region harbors a wealth of  young stars, emission nebulae, compact dark clouds and probable YSOs. Out of this material, we could derive the main properties of different populations located in the area.

First, we could assess that there are several populations along this particular line of sight: we could describe an spreading population of B-type stars with an important range in distance, two stellar concentrations ("a" and "b") located at $\sim$ 2.3 and $\sim$ 3.2~kpc, and some early stars ("c") apparently representing a young population located as distant as 8~kpc.

Second, by focusing our attention on the main stellar concentrations, we revealed that they are hosting a relevant population of PMS stars and are as young as 2-3~Myrs. We derived their IMFs and we found their slope for massive stars is low but still compatible with the standard Salpeter law, and we also provided an estimate of each stellar group in $\sim 1100$ and $\sim 500 M_{\odot}$.

Third, we found that all O stars in these two groups share similar kinematics and their representative values are different from that expected from Galactic rotation models in this part of the Milky Way. Regarding the mentioned young background population ("c"), we found it could had an apparent different kinematics we estimated an age of $\sim$ 10~Myr as a crude approximation.

Finally, we compared all these populations locations and characteristic with those results obtained at similar directions of the Galaxy and their are consistent with the idea of an almost tangent view along of the Sagittarius-Carina spiral arm. In particular, groups "a" and "b" would coincide with the near side of this arm, while group "c" with the far side, and the spread of B stars a continuous young population characteristic of a grand design spiral structure.

\section*{Acknowledgments}

GB, MJR, MAC and JAP acknowledges support from CONICET (PIPs 112-201101-00301 and 112-201201-00226). The authors are much obliged for the use of the NASA Astrophysics Data System, of the $SIMBAD$ database and $ALADIN$ tools (Centre de Donn\'ees Stellaires --- Strasbourg, France), and of the WEBDA open cluster database. This publication also made use of data from:
a) the Two Micron All Sky Survey, which is a joint project of the University of Massachusetts and the Infrared Processing and Analysis Center/California Institute of Technology, funded by the National Aeronautics and Space Administration and the National Science Foundation;
b) the AAVSO Photometric All-Sky Survey (APASS), funded by the Robert Martin Ayers Sciences Fund.
c) the Midcourse Space Experiment (MSX). Processing of the data was funded by the Ballistic Missile Defense Organization with additional support from NASA Office of Space Science.
We thank R. Mart\'{\i}nez and H. Viturro for technical support.
Finally, we thank the referee, whose comments helped to improve significantly the paper.


\begin{thebibliography}{99}
\bibitem[\protect\citeauthoryear{Alter et al.}{1970}]{Alter_et_al_1970}
  Alter, G., Balazs, B., \& Ruprecht, J., eds. 1970, Catalogue of Star
  Clusters and Associations (2d ed. ; Budapest: Akad. Kiado)

\bibitem[\protect\citeauthoryear{Allen}{1973}]{Allen_1973}
  Allen, C. W. 1973. Astrophysical Quantities, Third Edition.
  University of London, The Athlone Press.

\bibitem[\protect\citeauthoryear{Anderson \& Francis}{2012}]{Anderson_Francis_2012}
  Anderson, E., \& Francis, C.\ 2012, Astronomy Letters, 38, 331

\bibitem[\protect\citeauthoryear{Ardeberg \& Maurice}{1977}]{Ardeberg_Maurice_1977}
  Ardeberg, A., \& Maurice, E.\ 1977, A\&AS, 28, 153

\bibitem[\protect\citeauthoryear{Ardeberg \& Maurice}{1980}]{Ardeberg_Maurice_1980}
  Ardeberg, A., \& Maurice, E.\ 1980, A\&AS, 39, 325

\bibitem[\protect\citeauthoryear{Ardeberg \& Maurice}{1981}]{Ardeberg_Maurice_1981}
  Ardeberg, A., \& Maurice, E.\ 1981, A\&A, 98, 9

\bibitem[\protect\citeauthoryear{Aumer \& Binney}{2009}]{Aumer_Binney_2009}
  Aumer, M., \& Binney, J.~J.\ 2009, MNRAS, 397, 1286

\bibitem[\protect\citeauthoryear{Baume et al.}{2003}]{Baume_et_al_2003}
  Baume, G., V{\'a}zquez, R.~A., Carraro, G., \& Feinstein, A.\ 2003, A\&A, 402, 549

\bibitem[\protect\citeauthoryear{Baume et al.}{2009}]{Baume_et_al_2009}
  Baume, G., Carraro, G., \& Momany, Y.\ 2009, MNRAS, 398, 221

\bibitem[\protect\citeauthoryear{Baume et al.}{2011}]{Baume_et_al_2011}
  Baume, G., Carraro, G., Comeron, F., \& de El{\'{\i}}a, G.~C.\ 2011, A\&A, 531, A73

\bibitem[\protect\citeauthoryear{Borissova et al.}{2011}]{Borissova_et_al_2011}
  Borissova, J., Bonatto, C., Kurtev, R., et al.\ 2011, A\&A, 532, A131



 \bibitem[\protect\citeauthoryear{Carraro}{2013}]{Carraro 2013}
  Carraro, G., 2013,  Proceedings of the IAU Symposium No. 298,
  "Setting the scene for Gaia and LAMOST",
  eds. S. Feltzing, G. Zhao, N. A. Walton, and P. A. Whitelock, in press ({\tt arXiv:1307.0569})

\bibitem[\protect\citeauthoryear{Carraro \& Costa}{2009}]{Carraro_Costa_2009}
  Carraro, G., \& Costa, E.\ 2009, A\&A, 493, 71


\bibitem[\protect\citeauthoryear{Corti et al.}{2012}]{Corti_et_al_2012}
  Corti, M.~A., Arnal, E.~M., \& Orellana, R.~B.\ 2012, A\&A, 546, A62

\bibitem[\protect\citeauthoryear{Corti \& Orellana}{2013}]{Corti_Orellana_2013}
  Corti, M.~A., \& Orellana, R.~B.\ 2013, A\&A, 553, A108

\bibitem[\protect\citeauthoryear{Conti et al.}{1977}]{Conti_et_al._1977}
  Conti, P.~S., Leep, E.~M., \& Lorre, J.~J.\ 1977, ApJ, 214, 759

\bibitem[\protect\citeauthoryear{Cousins}{1978a}]{Cousins_1978a}
  Cousins, A.~W.~J.\ 1978, Monthly Notes of the Astronomical Society of South Africa, 37, 62

\bibitem[\protect\citeauthoryear{Cousins}{1978b}]{Cousins_1978b}
  Cousins, A.~W.~J.\ 1978, Monthly Notes of the Astronomical Society of South Africa, 37, 77 errata


\bibitem[\protect\citeauthoryear{Egan et al.}{2003}]{Egan_et_al_2003}
  Egan, M.~P., Price, S.~D., Kraemer, K.~E., et al.\ 2003, VizieR Online Data Catalog, 5114, 0

\bibitem[\protect\citeauthoryear{Fich et al.}{1989}]{Fich_et_al_1989}
  Fich, M., Blitz, L., \& Stark, A.~A.\ 1989, ApJ, 342, 272

\bibitem[\protect\citeauthoryear{Feigelson \& Getman}{2005}]{Feigelson_Getman_2005}
  Feigelson, E.~D., \& Getman, K.~V.\ 2005, arXiv:astro-ph/0501207

\bibitem[\protect\citeauthoryear{Fruchter \& Hook}{2002}]{Fruchter_Hook_2002}
  Fruchter, A.~S., \& Hook, R.~N.\ 2002, PASP, 114, 144

\bibitem[\protect\citeauthoryear{Gallart et al.}{2003}]{Gallart_et_al_2003}
  Gallart, C., Zoccali, M., Bertelli, G., et al.\ 2003, AJ, 125, 742

\bibitem[\protect\citeauthoryear{Gies et al.}{2002}]{Gies_et_al_2002}
  Gies, D.~R., Penny, L.~R., Mayer, P., Drechsel, H., \& Lorenz, R.\ 2002, ApJ, 574, 957



\bibitem[\protect\citeauthoryear{Henden et al.}{2010}]{Henden_et_al_2010}
  Henden, A.~A., Terrell, D., Welch, D., \& Smith, T.~C.\ 2010,
  Bulletin of the American Astronomical Society, 42, 515

\bibitem[\protect\citeauthoryear{Hou et al.}{2009}]{Hou_et_al_2009}
  Hou, L.~G., Han, J.~L., \& Shi, W.~B.\ 2009, A\&A, 499, 473

\bibitem[\protect\citeauthoryear{Huang \& Gies}{2006}]{Huang_Gies_2006}
  Huang, W., \& Gies, D.~R.\ 2006, ApJ, 648, 580

\bibitem[\protect\citeauthoryear{Jester et al.}{2005}]{Jester_et_al_2005}
  Jester, S., Schneider, D.~P., Richards, G.~T., et al.\ 2005, AJ, 130, 873

\bibitem[\protect\citeauthoryear{Kenyon \& Hartmann}{1990}]{Kenyon_Hartmann_1990}
  Kenyon, S.~J., \& Hartmann, L.~W.\ 1990, ApJ, 349, 197

\bibitem[\protect\citeauthoryear{Kharchenko et al.}{2007}]{Kharchenko_et_al_2007}
  Kharchenko, N.~V., Scholz, R.-D., Piskunov, A.~E., R{\"o}ser, S.,
  \& Schilbach, E.\ 2007, Astronomische Nachrichten, 328, 889

\bibitem[\protect\citeauthoryear{Kharchenko \& Roeser}{2009}]{Kharchenko_Roeser_2009}
  Kharchenko, N.~V., \& Roeser, S.\ 2009, VizieR Online Data Catalog, 1280, 0

\bibitem[\protect\citeauthoryear{Kippenhahn et al.}{1967}]{Kippenhahn_1967}
  Kippenhahn, R., Weigert, A., Hofmeister, E., 1967, ``Methods for Computational Physics'',
  Vol. 7 in Alder B., P.129 New York Academic Press

\bibitem[\protect\citeauthoryear{Koornneef}{1983}]{Koornneef_1983}
  Koornneef, J.\ 1983, A\&A, 128, 84

\bibitem[\protect\citeauthoryear{Lada and Lada}{2003}]{Lada_Lada 2003}
  Lada, C. J. \& Lada, E. A., 2003, ARA\&A, 41, 57

\bibitem[\protect\citeauthoryear{Lumsden et al.}{2002}]{Lumsden_et_al_2002}
  Lumsden, S.~L., Hoare, M.~G., Oudmaijer, R.~D., \& Richards, D.\ 2002, MNRAS, 336, 621

\bibitem[\protect\citeauthoryear{Majaess et al.}{2009}]{Majaess_et_al_2009}
  Majaess, D.~J., Turner, D.~G., \& Lane, D.~J.\ 2009, MNRAS, 398, 263

\bibitem[\protect\citeauthoryear{Marigo et al.}{2008}]{Marigo_et_al_2008}
  Marigo, P., Girardi, L., Bressan, A., et al.\ 2008, A\&A, 482, 883

\bibitem[\protect\citeauthoryear{Martins \& Plez}{2006}]{Martins_Plez_2006}
  Martins, F., \& Plez, B.\ 2006, A\&A, 457, 637

\bibitem[\protect\citeauthoryear{McSwain \& Gies}{2005}]{McSwain_Gies_2005}
  McSwain, M.~V., \& Gies, D.~R.\ 2005, ApJS, 161, 118

\bibitem[\protect\citeauthoryear{Naz{\'e} et al.}{2013}]{Naze_et_al_2013}
  Naz{\'e}, Y., Rauw, G., Sana, H., \& Corcoran, M.~F.\ 2013, A\&A, 555, A83

\bibitem[\protect\citeauthoryear{Noriega-Crespo et al.}{1997}]{Noriega-Crespo_et_al_1997}
  Noriega-Crespo, A., van Buren, D., \& Dgani, R.\ 1997, AJ, 113, 780

\bibitem[\protect\citeauthoryear{Peri et al.}{2012}]{Peri_et_al_2012}
  Peri, C.~S., Benaglia, P., Brookes, D.~P., Stevens, I.~R., \& Isequilla, N.~L.\ 2012, A\&A, 538, A108

\bibitem[\protect\citeauthoryear{Perry \& Landolt}{1986}]{Perry_Landolt_1986}
  Perry, C.~L., \& Landolt, A.~U.\ 1986, AJ, 92, 844


\bibitem[\protect\citeauthoryear{Preibisch \& Zinnecker}{1999}]{Preibisch_Zinnecker_1999}
  Preibisch, T., \& Zinnecker, H.\ 1999, AJ, 117, 2381

\bibitem[\protect\citeauthoryear{Prisinzano et al.}{2005}] {Prisinzano_2005}
  Prisinzano, L., Damiani, F., Micela, G.,
  \& Sciortino, S.\ 2005, A\&A, 430, 941

\bibitem[\protect\citeauthoryear{Reipurth et al.}{1997}]{Reipurth_et_al_1997}
  Reipurth, B., Corporon, P., Olberg, M., \& Tenorio-Tagle, G.\ 1997, A\&A, 327, 1185

\bibitem[\protect\citeauthoryear{Reipurth}{2008}]{Reipurth_2008}
  Reipurth, B.\ 2008, Handbook of Star Forming Regions, Vol. II

\bibitem[\protect\citeauthoryear{Salpeter}{1955}]{Salpeter_1955}
  Salpeter, E.~E.\ 1955, ApJ, 121, 161

\bibitem[\protect\citeauthoryear{Sana et al.}{2011}]{Sana_et_al_2011}
  Sana, H., James, G., \& Gosset, E.\ 2011, MNRAS, 416, 817

\bibitem[\protect\citeauthoryear{Schmidt-Kaler}{1982}]{Schmidt-Kaler_1982}
  Schmidt-Kaler, Th. 1982, Landolt-B\"ornstein, Numerical data and
  Functional Relationships in Science and Technology, New Series,
  Group VI, Vol. 2(b), ed. K. Schaifers, \& H. H. Voigt (Berlin:
  Springer Verlag), 14

\bibitem[\protect\citeauthoryear{Siess et al.}{2000}]{Siess_et_al_2000}
  Siess, L., Dufour, E., \& Forestini, M.\ 2000, A\&A, 358, 593

\bibitem[\protect\citeauthoryear{Skrutskie et al.}{2006}]{Skrutskie_et_al_2006}
  Skrutskie, M.~F., Cutri, R.~M., Stiening, R., et al.\ 2006, AJ, 131, 1163

\bibitem[\protect\citeauthoryear{Sota et al.}{2013}]{Sota_et_al_2013}
  Sota, A., Ma{\'{\i}}z Apell{\'a}niz, J., Morrell, N.~I., et al.\ 2013, arXiv:1312.6222

\bibitem[\protect\citeauthoryear{Stetson}{1987}]{Stetson_1987}
  Stetson, P.~B.\ 1987, PASP, 99, 191

\bibitem[\protect\citeauthoryear{Stetson}{1992}]{Stetson_1992}
  Stetson, P.~B.\ 1992, in Stellar Photometry-Current Techniques and
  Future Developments, ed. C. J. Bulter, \& I. Elliot (Cambridge:
  Cambridge University Press), IAU Coll., 136, 291


\bibitem[\protect\citeauthoryear{Thackeray \& Wesselink}{1965}]{Thackeray_Wesselink_1965}
  Thackeray, A.~D., \& Wesselink, A.~J.\ 1965, MNRAS, 131, 121

\bibitem[\protect\citeauthoryear{Vall{\'e}e}{2008}]{Vallee_2008}
  Vall{\'e}e, J.~P.\ 2008, AJ, 135, 1301

\bibitem[\protect\citeauthoryear{van Buren et al.}{1995}]{van_Buren_1995}
  van Buren, D., Noriega-Crespo, A., \& Dgani, R.\ 1995, AJ, 110, 2914


\bibitem[\protect\citeauthoryear{Vega et al.}{1994}]{Vega_et_al_1994}
  Vega, E.~I., Orsatti, A.~M., \& Marraco, H.~G.\ 1994, AJ, 108, 1834

\bibitem[\protect\citeauthoryear{Walborn}{1972}]{Walborn_1972}
  Walborn, N.~R.\ 1972, AJ, 77, 312

\bibitem[\protect\citeauthoryear{Walborn}{1987}]{Walborn_1987}
  Walborn, N.~R.\ 1987, AJ, 93, 868


\bibitem[\protect\citeauthoryear{Zacharias et al.}{2013}]{Zacharias_et_al_2013}
  Zacharias, N., Finch, C.~T., Girard, T.~M., et al.\ 2013, AJ, 145, 44

\bibitem[\protect\citeauthoryear{Zwicky}{1957}]{Zwicky_1957}
  Zwicky, F.\ 1957, PASP, 69, 518

\end{thebibliography}
\end{document}